# Experimental studies of 7-cell dual axis asymmetric cavity for energy recovery linac

I. V. Konoplev,[1,*] K. Metodiev,[1] A. J. Lancaster,[1] G. Burt,[2] R. Ainsworth,[3] and A. Seryi[1]

[1]*JAI, Department of Physics, University of Oxford, Oxford OX1 3RH, United Kingdom*
[2]*Cockcroft Institute, Lancaster University, Lancaster LA1 4YW, United Kingdom*
[3]*Fermilab, Batavia, Illinois 60510, USA*

High average current, transportable energy recovery linacs (ERLs) can be very attractive tools for a number of applications including next generation high-luminosity, compact light sources. Conventional ERLs are based on an electron beam circulating through the same set of rf cavity cells. This leads to an accumulation of high-order modes inside the cavity cells, resulting in the development of a beam breakup (BBU) instability, unless the beam current is kept below the BBU start current. This limits the maximum current which can be transported through the ERL and hence the intensity of the photon beam generated. It has recently been proposed that splitting the accelerating and decelerating stages, tuning them separately and coupling them via a resonance coupler can increase the BBU start current. The paper presents the first experimental rf studies of a dual axis 7-cell asymmetric cavity and confirms the properties predicted by the theoretical model. The field structures of the symmetric and asymmetric modes are measured and good agreement with the numerical predictions is demonstrated. The operating mode field flatness was also measured and discussed. A novel approach based on the coupled mode (Fano-like) model has been developed for the description of the cavity eigenmode spectrum and good agreement between analytical theory, numerical predictions and experimental data is shown. Numerical and experimental results observed are analyzed, discussed and a good agreement between theory and experiment is demonstrated.

## I. INTRODUCTION

Transportable sources capable of efficient generation of high luminosity and intensity photon beams in THz, EUV and x-ray regions are attractive tools for a number of applications [1–6]. Today such beams are generated mainly at large scale, national facilities such as free electron lasers (FEL) and synchrotron radiation sources (SRC). In the context of this paper we will refer to a source as transportable, if its footprint is around 10 m$^2$ and it can be transported using a conventional trailer (for example) as opposed to the national facilities, which are large, expensive and exclusive to a limited number of high level research activities. Providing photon beams of a similar quality to a broader scientific community could potentially generate ground breaking results in many branches of science and industry including biology, chemistry, pharmacology, medical science, security, etc. The limited availability of such instruments is hindering research and development. One way to resolve this issue is to create next generation light sources based on an energy recovery linac (ERL), which will complement the current photon factories. The ERLs are driven by electron beams with low average current which means a relatively low average intensity of the radiation. The current limitations linked to the development of a beam breakup (BBU) instability [7,8], which develops in conventional single axis SCRF ERLs or strongly coupled dual axis systems [9,10] if the beam current is increased above some threshold value which is usually around 100 mA. The dual axis asymmetric energy recovery linac (AERL) in which acceleration and deceleration are separated, while the cells are individually tuned and linked via a resonant coupler has recently been suggested [11–14] to mitigate the development of the instability. The theoretical studies of an 11-cell cavity [11] indicated the possibility to increase the BBU starting current, thus stimulating the research.

The aim of this paper is to validate the previous results using experimental and numerical approaches, develop new experimental techniques and data analysis, and to demonstrate the scalability of the concept developed. The outcome of the first experimental studies of a dual axis asymmetric cavity (Fig. 1) will be presented and discussed. To demonstrate scalability of the conceptual design i.e. conservation of the basic properties predicted in [11] with variation of the cavity length, to speed up the cavity manufacturing and save the cost of machining a 7-cell

*Corresponding author.
ivan.konoplev@physics.ox.ac.uk

## II. THE 7-CELL CAVITY NUMERICAL STUDIES AND EXPERIMENTAL SETUP

Here the 7-cell dual axis asymmetric cavity is studied using both numerical and experimental techniques in order to find its spectral characteristics (eigenmode $Q$-factors, their frequency position and measuring the fields' profiles). In Fig. 1 the drawing of the cavity generated by CST Microwave Studio is shown and the ports numbers, which will be used throughout this paper, are indicated. The dimensions of the 7-cell and the 11-cell cavities for a specific cell located on the same axis are identical except that the 11-cell structure has an increased number of middle cells on each axis (3 instead of 1). The scaling from 11-cell to 7-cell was done to demonstrate the scalability of the model, which could be required for instance to reduce the cost of the cavity manufacturing. To cover a broad range of the modes, cavity (Fig. 1) was considered for the first prototype studies. The 7-cell rf cavity high-$Q$ modes in frequency range from 1 to 2.5 GHz are studied and the results are presented. All experimental data will be compared with theoretical predictions and discussed. The properties of the eigenmodes including their coupling are examined, and different models are used to describe them. The modes were experimentally identified, studied and their field structures were measured.

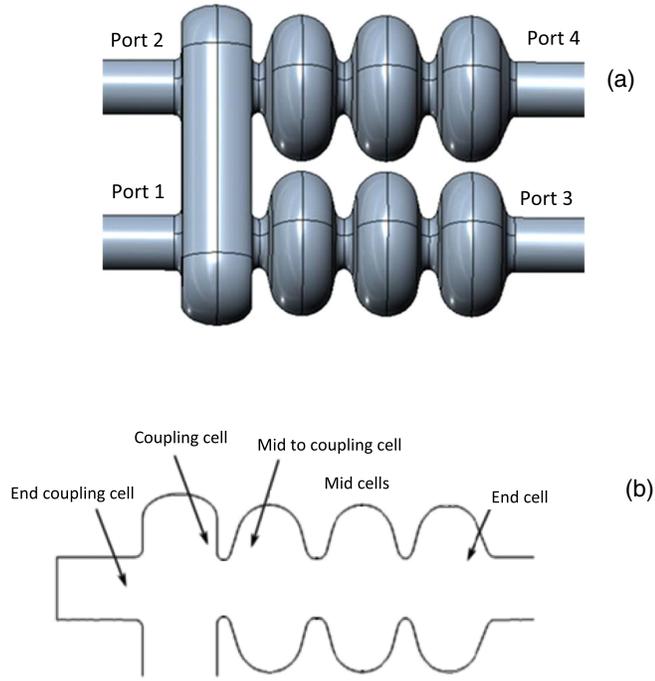

FIG. 1. (a) The 3D view of the 7-cell cavity generated by the CAD software and used in 3D CST Microwave Studio. The port numbering is used throughout the text as shown. (b) Part of the technical drawing indicating the cavity cells similar to models studied in [11].

The outline of the rest of the paper is as follows. In the second section the results of numerical studies of a 7-cell cavity using CST Microwave (CST MW) Studio will be presented and discussed. The third section will be dedicated to the cavity modes' $Q$-factor measurements. In the fourth section the experimental results of the eigenmodes' field structure measurements will be presented and compared with numerical predictions. In the conclusion we will discuss further steps to realize asymmetric energy recovery Linac (AERL). We note that the numerical studies were carried out using a 3D CST MW Studio software package, instead of ACE3P electromagnetic suite (developed at SLAC) and used in the previous work [11]. The new software was used to check validity of the conceptual model and compare the capabilities of the broadly available commercial product 3D CST Microwave Studio with highly reputable specialist software ACE3P. The numerical results observed from two different codes were similar further confirming the results observed.

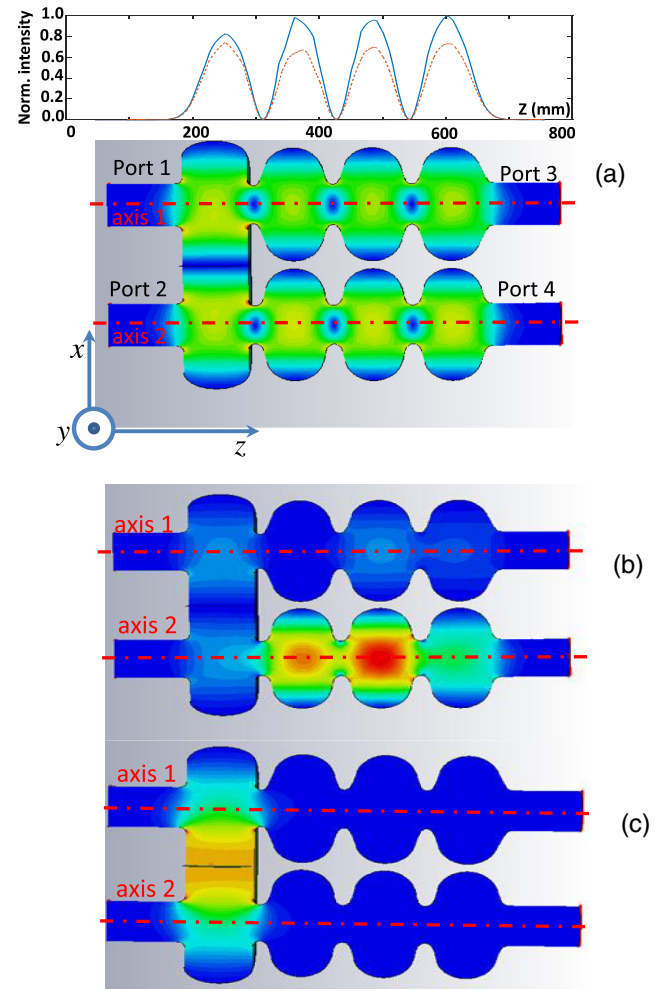

FIG. 2. The contour plots of the electric fields of eigenmodes calculated by the full 3D code CST Microwave Studio and showing (a) symmetric mode, (b) asymmetric mode and (c) bridge mode. The ports and center lines (dash-dotted lines) are shown. The inset of (a) shows the comparison of the intensities of the electric fields on the axis, field flatness is around 85%.

the studies were conducted in the frequency range from 1 to 2.5 GHz. The cavity was machined using the technical CAD drawing developed for the numerical studies.

The numerical investigations conducted show (similar to [11]) that the cavity eigemodes can be separated into symmetric and asymmetric eigenmodes, and eigenmodes associated with the resonant coupler cell. In Figs. 2(a)–2(c) examples of the contour plots of the electric field distribution of these eigenmodes are shown. The inset in Fig. 2(a) (above the figure) shows the dependence of the normalized intensity of the electric field along both axes. A field flatness of around 85% was observed. As in previous studies [11] the eigenmodes associated with the resonant coupler [Fig. 2(c)] have similar structures and are localized inside the coupling cell and the contour plots are similar to those observed before, for the 11-cell cavity. In Fig. 3(a) the cavity eigenmode spectrum in the frequency range 1 to 2.5 GHz is shown. There are "bridge modes" at 1.099 and 1.495 GHz which are associated with the resonant coupler [Fig. 2(c)], passband symmetric and asymmetric modes in the interval 1.27 to 1.3 GHz and high order modes located above 1.5 GHz. In this paper only the passband and bridge modes are considered, while a detailed investigation of HOM is outside the scope of this paper. The number of eigenmodes in the passband interval is 7 (as expected from simulations) and a close-up of these modes is shown in the inset of Fig. 3(a). In Fig. 3(b) the graphs show the dependences of the normalized intensities of the electric fields along the cavity for the bridge mode located at 1.099 GHz (dotted line), operating mode (1.299 229 GHz) and nearest high Q asymmetric mode (dashed line) located at 1.289 298 GHz. The field distributions of asymmetric modes and symmetric modes are different, which can be used in the tuning of the cavity to improve its performance, however more detailed field analysis will be required to suppress the asymmetric mode.

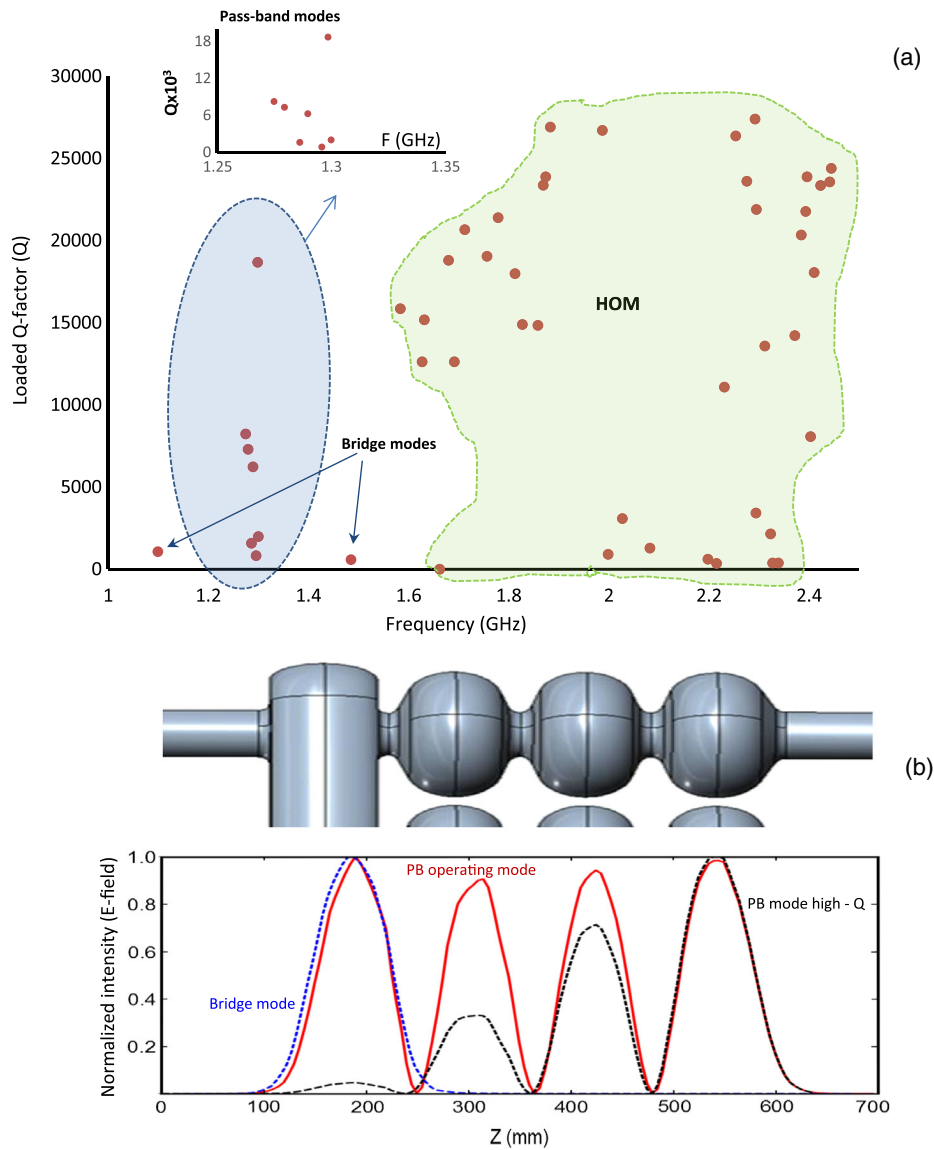

FIG. 3. (a) The 7-cell cavity eigenmodes spectrum (1–2.5 GHz) calculated by the 3D CST MW Studio with bridge modes, passband (PB) modes and high order modes (HOM) indicated. The inset of the figure is a zoom-in (1.25–1.35 GHz) of the spectra showing all seven passband modes. (b) Calculated normalized field intensity versus the longitudinal coordinates (overlaid over the 3D plot of the cavity for visualization purpose) for the bridge mode (dotted line), PB operating mode (solid line) and the passband mode closest to the operating mode (dashed line).

To verify the results of the numerical studies a dual axis 7-cell rf cavity (Fig. 1) has been machined from aluminum and the experimental investigations have been carried out.

A photograph of the aluminum cavity is shown in Figs. 4(a) and 4(b). It was built from two solid blocks of aluminum [Fig. 4(a)] using a computer controlled milling machine, the internal surfaces of the machined blocks were polished and cleaned prior to its assembly and positioning on the rf test table. Machining the cavity in this way allows for quick assembly by connecting the two blocks and securing them with pins and screws [Fig. 4(b)], as well as quick disassembly if required. There is no soldering involved, which makes the whole system flexible for cavity tuning and rf studies of new prototype. After the cavity was assembled, it was checked for imperfections and discontinuities (potentially caused by misalignment) along the internal surfaces. No obvious discontinuities inside were detected and the seams were shown to be good enough to consider the joints to be seamless. Clearly, there are a number of limitations associated with such machining and any TESLA like SCRF cavity will be built using the conventional pressing and electron beam welding technique [15]. Discussion of these limitations is outside the scope of this paper, however, the prototype is relatively inexpensive and convenient for basic studies of rf properties. The photograph [Fig. 4(a)] shows the distinctive "$3 + 3 + 1$" configuration where six cells ($3 + 3$) (each similar to a conventional 1.3 GHz TESLA cavity) are linked ($+1$) by a resonant coupler, which can be easily distinguished. In Fig. 4(b) a photograph of the rf test table with the Vector Network Analyser (VNA) and the cavity assembled is shown. The two-port VNA (Anritsu MS4644B) was used to perform measurements and the pillars visible on the figure are part of the rf bead-pull setup [16]. The VNA was also used to carry out studies of the cavity spectrum. All the measurements conducted were very sensitive to the external environment including humidity, temperature variation, acoustic noise level and the mechanical vibrations of the ground. As a result some of the measurements were carried out during the night to improve the results and special air stoppers were designed and installed to prevent any airflow inside the cavity. As an illustration of conventional machining Fig. 4(c) shows the next step of the dual axis cavity studies. The constituent parts of a copper 11-cell cavity were manufactured using a pressing technique, opposite to the technique discussed above. To test conventional manufacturing techniques (pressing, cleaning, welding and securing the position of the cells), the parts shown were built using copper sheets which have similar mechanical properties to niobium. The cavity will be cleaned and fully assembled using a conventional welding technique and the arms will be secured to assure that the axes are parallel and are on the same plane.

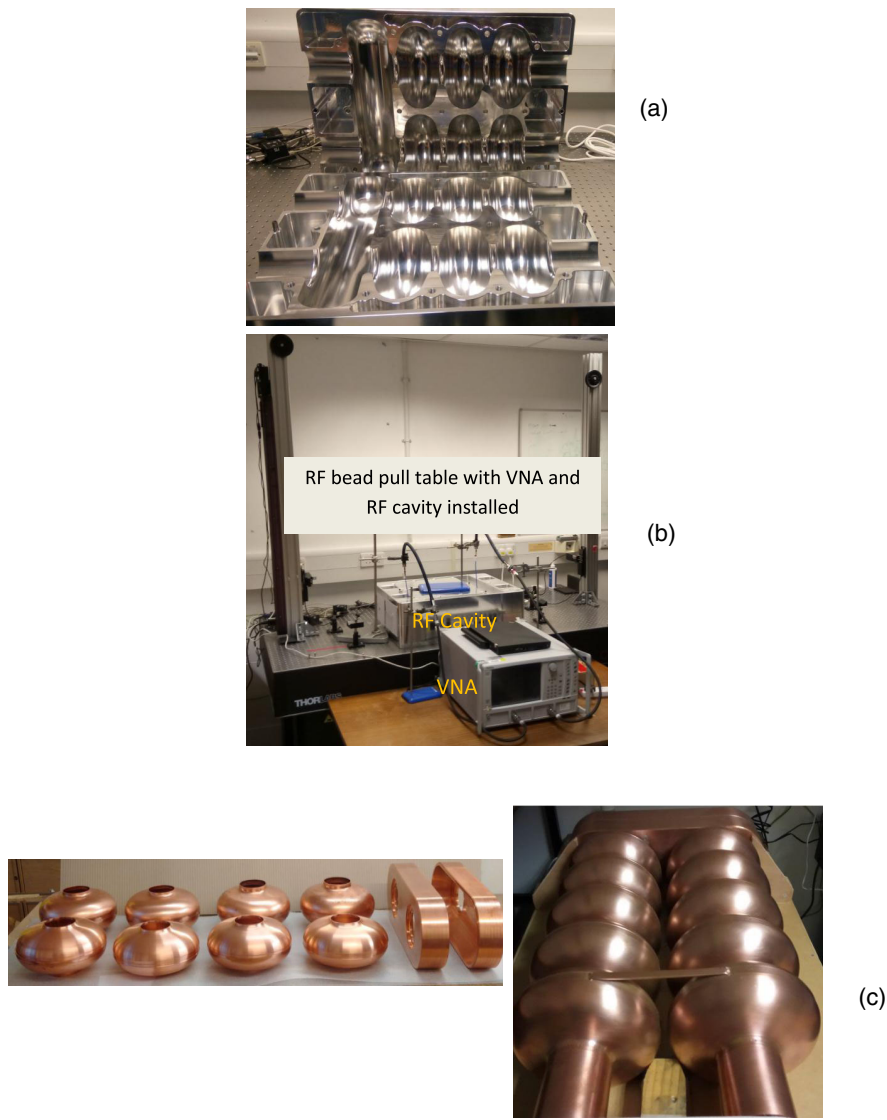

FIG. 4. (a) The photograph of the 7-cell aluminum cavity before assembly machined from two blocks of aluminum, (b) the assembled 7-cell cavity on the rf test bench, (c) separate parts of the copper 11-cell cavity machined using the conventional sheet-press technology which would be used to manufacture a SC rf cavity (left photograph) and fully assembled 11-cell cavity (right photograph).

## III. EXPERIMENTAL MEASUREMENTS OF 7-CELLS RF CAVITY SPECTRUM

An experimental study of the spectrum of a dual axis cavity has been carried out and compared with theoretical predictions. To the best knowledge of the authors there have not been experimental studies of such a two axis system conducted and published previously. Therefore the advantages and disadvantages of different techniques [17–19] developed to carry out rf studies of conventional single axis cavities were not clear with respect to the new dual axis system. Two techniques were tested to evaluate the eigenmodes' $Q$-factors: the first is based on the "reflection" $S_{ii}$ measurements and second on the "transmission" $S_{ij}$ measurements ($i$ and $j$ represent the port numbers, as shown in Fig. 1). In Fig. 5 the schematics illustrating the experimental setup are shown. During the measurements the electric dipole couplers were inserted along the same axis at opposite ports (Fig. 5). The reflection measurements [17] are associated with the measurements of the complex $S_{ii}$ parameter and eigenmode reconstruction from the $S_{11}$ polar diagram. To do this, a free space measurement [Fig. 6(a)] was made, to calibrate and evaluate perturbations of the cavity polar diagram [Fig. 6(b)] consequently recovering the eigenmodes' positions and their $Q$-factor. The small perturbations on Fig. 6(b) show the eigenmodes and, using the techniques described in [17], the spectrum of these eigenmodes can be constructed. However, due to weak coupling and high sensitivity to external factors, the uncertainty level of the measurements was very high and the method was abandoned in favor of the second technique.

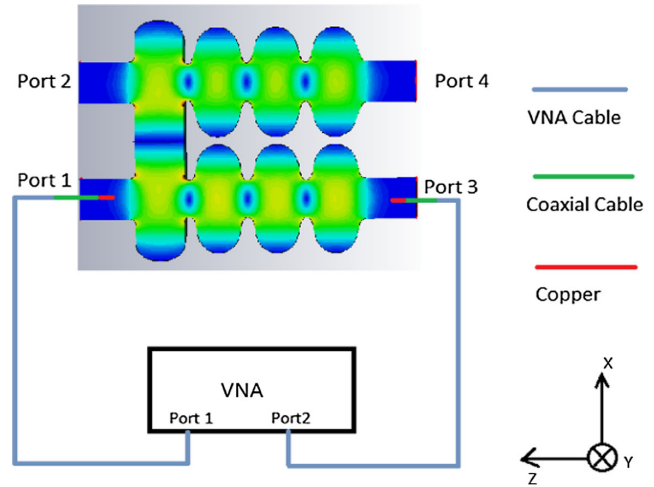

FIG. 5. Schematic of the experimental setup to measure the transmission $S_{ij}$ parameter.

The second technique was based on transmission $S_{ij}$ measurements and the recovery of the eigenmode spectrum directly from measured data. Initially there were concerns about radiation losses from the "free" ports, for example ports 1–3 and 2–4, if $S_{24,13}$ are measured respectively. Also the effect of the couplers on the eigenmodes loaded $Q$-factors was unclear. However, a metal plate located on the outside close to the open port and thus effectively blocking the port resulted in a relatively small phase difference of the measured $S_{21}$ parameter (below 0.1°), while simple mechanical vibrations of the apparatus could lead to at least of order ~1° phase variations. A similar negligible effect due to the couplers was also observed. In Fig. 7 typical $S_{ij}$ characteristics (solid line) in the frequency range 1.27–1.3 GHz is shown. During the study deviations over the course of a measurement were seen to be insignificant. This allowed us to carry out the measurements assuming a negligible influence from couplers and open ports. There are seven well defined, visible peaks (Fig. 7, solid line) and each peak corresponds to an eigenmode with a specific frequency position and $Q$-factor, which coincide well with the numerical predictions [Fig. 3(a)].

Prior to any discussion of the results (Fig. 7, bold line), it is important to consider the eigenmodes of the cavity.

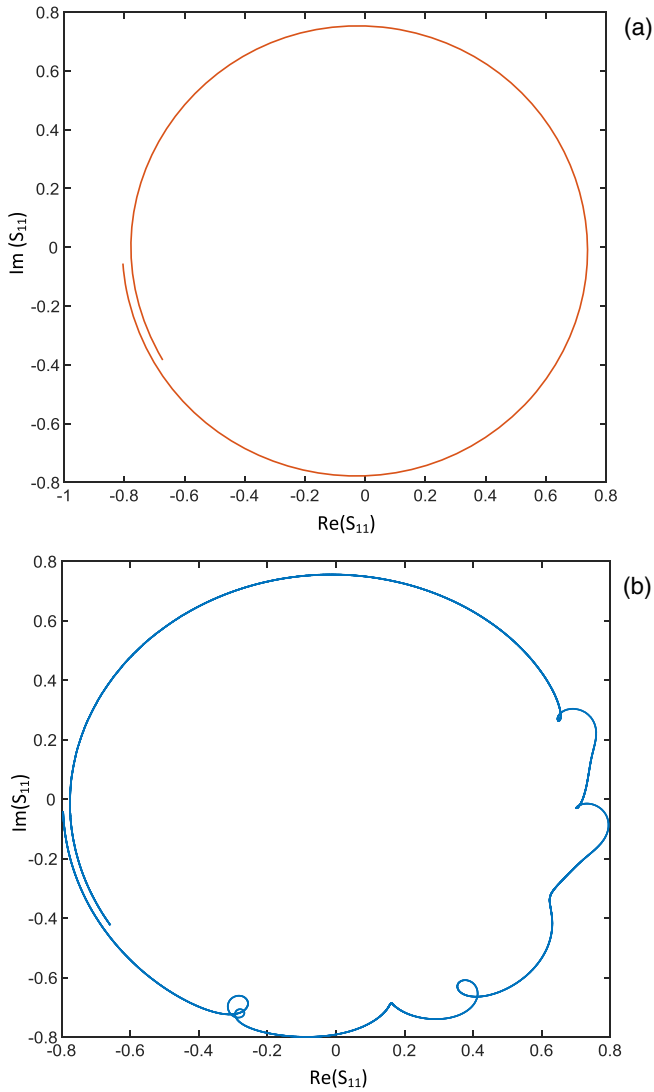

respectively. The solution of the equation (assuming a wave solution $\sim e^{i\omega t}$) gives the following function as the best fit for the spectral line of the eigenmode:

$$|A| = \frac{\text{const}}{\sqrt{(\omega - \omega_0)^2 + \omega^2 \gamma^2}} \quad (2)$$

which in the approximation $\omega \approx \omega_0$ leads (using a Taylor expansion) to the well-defined and understood shape (Lorentzian) of the eigenmode spectral line in the vicinity of the resonance frequency $|A| \approx \frac{\text{const}}{\sqrt{1+4(\frac{\omega-\omega_0}{\gamma})^2}}$. To define the parameters of the eigenmodes using the experimental data observed (Fig. 7, bold line) and to construct a theoretical $S_{ij}$ curve which fits experimental data, the solutions defined by (2) can be used. The $S_{ij}$ curve can be presented as a superposition of the amplitudes [18–20]: $S_{12} = \sum A_i(\omega_0^i, \gamma_i, \omega)$; which takes into account small background interference (observed) and considering only a small frequency range (inside which the fitting is done) has the following form:

$$S_{21} = A_1 + A_2 f + \sum_{i=1}^{N} |S_{21\,\text{max}}^i| / \sqrt{1 + 4(\frac{(\omega - \omega_0^i)}{\gamma_i})^2}, \quad (3)$$

where $N$ is the number of maxima measured and the loaded $Q$-factor is defined as $Q_i = \frac{\omega_0^i}{\gamma_i}$. Using this model, a theoretical curve $S_{ij}$ was built, and the results are shown in Fig. 7 (dotted line) with the positions of the eigenmodes and their loaded $Q$-factors (used to calculate the theoretical

FIG. 6. The $S_{11}$ polar diagrams in frequency range (1.27–1.305 GHz) of (a) free space (background measurements used for calibration purposes) and (b) the cavity. The small perturbations seen in (b) are due to excitation of the cavity eigenmodes and each deviation from the free space measurements (a) indicates a possible eigenmode of the cavity.

The eigenmodes of a conventional single axis rf cavity can be described using an RLC circuit approach with the loaded $Q$-factor $Q = 2\pi W_{st}/W_d$, i.e. ratio between stored energy $W_{st}$ in the cavity and dissipated (including radiation) energy $W_d$ from the cavity per oscillating cycle. The function which defines an eigenmode spectral line can also be found analytically from the dumped harmonic oscillator equation:

$$\ddot{x} + \gamma \dot{x} + \omega_0^2 x = F_0 e^{i\omega t}, \quad (1)$$

where $x$ is the amplitude of oscillations (here the field amplitude), $\gamma$, $\omega_0$ and $F_0$ are the damping parameter, eigenfrequency of oscillator and external harmonic force,

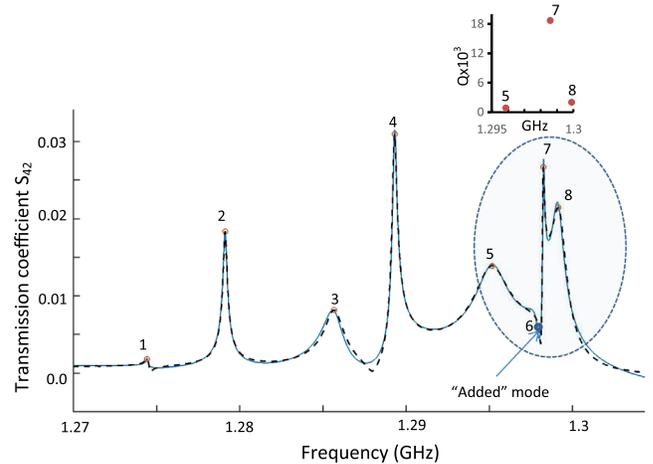

FIG. 7. The graph for the frequency interval (1.27 to 1.305 GHz) shows a comparison of semianalytical predictions based on an uncoupled oscillator model (dashed line) with $S_{42}$ measurements (solid line) and CST Microwave Studio prediction of the eigenmodes spectral positions (empty circles). The solid circle indicates an additional mode needed to fit the semianalytical model with the measurements. The inset zooms in on the modes of interest as predicted by CST MW Studio.

predictions) illustrated by circles. However, in order to find and build this theoretical curve, the parameters $\omega_0^i$, $\gamma_i$ and the number of eigenmodes $N$ have to be identified using a numerical approach. An iterative approach was used to find the best fit theoretical line (requiring the trivial computing resources) with all eigenmode parameters identified, including the frequencies of the eigenmode $\omega_0^i$ (dots) and their loaded $Q_i$-factor. However, the theoretical line (Fig. 7) shows only partial agreement with measurements. There are a number of inaccuracies including an appearance of an additional erroneous eigenmode (number 6, solid dot) which was not measured. Taking into account the results and the precision of the measurements (frequency step is below 1.5 KHz) it was expected that $N$ should be equal to the number of the peaks measured and the parameters $\omega_0^i$, $\gamma_i$ should also coincide well with the measured values. An introduction of an additional mode led to the realization that the approach based on the uncoupled modes assumption [Eq. (1)] does not provide an accurate solution. The inset of Fig. 7 illustrates the set of eigenmodes calculated numerically using CST Microwave Studio which is different from the one observed using the uncoupled model. Therefore a coupled oscillators model [21–24] with Fano-like solutions was adopted. The mode coupling could take place for a number of different reasons including the presence of a resonant coupling cell and the fact that each cell is slightly detuned from its neighbor. The mechanism of coupling and study of this complex phenomena inside the cavity is outside the scope of this paper and will be considered in future work. In the case of the

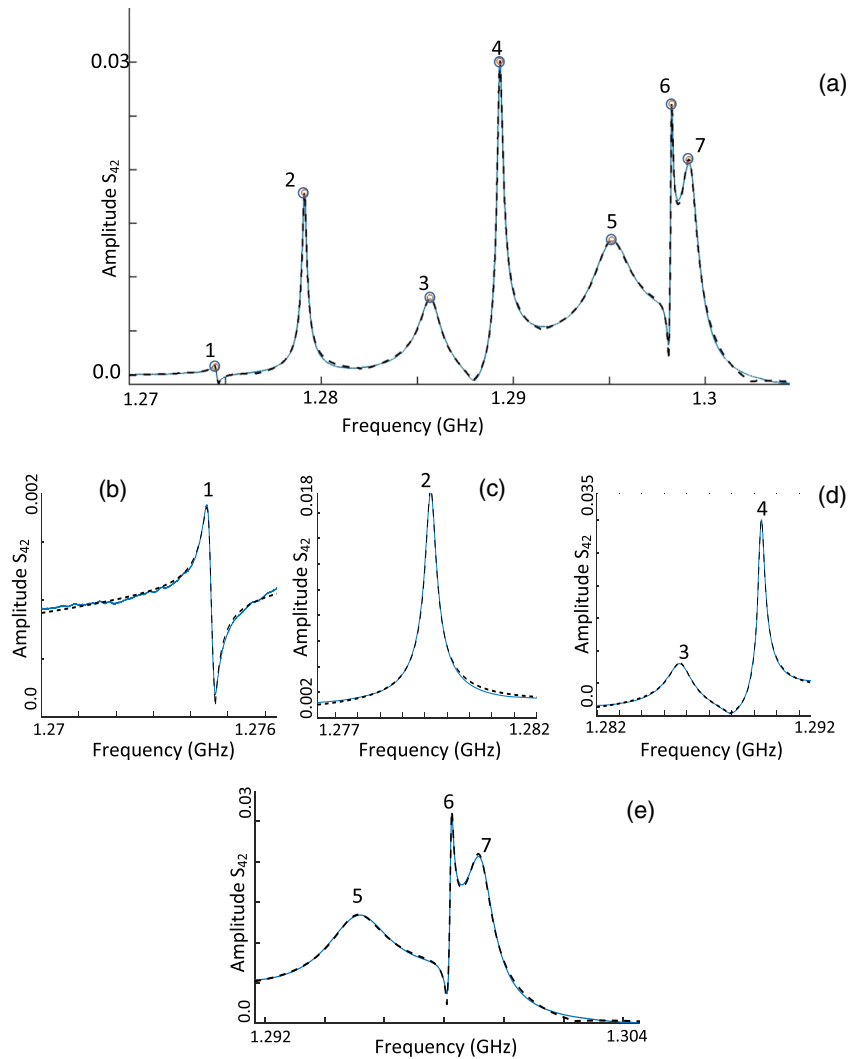

FIG. 8. (a) Comparison of measured (blue solid line) and calculated (black dashed line) transmission coefficients $S_{42}$. The calculated curve was observed using a semianalytical model based on the coupled oscillators assumption (Fano model) in the frequency interval (1.27–1.305 GHz). The empty circles indicate (as in Fig. 7) the eigenmode positions calculated by CST MW Studio. Comparison of measured (solid line) and calculated (dashed line) transmission coefficients $S_{42}$ using the Fano model in frequency intervals: (b) 1.27–1.276 GHz; (c) 1.276–1.282 GHz; (d) 1.282–1.292 GHz; (e) 1.292–1.304 GHz.

coupled oscillators model the modes can be defined by the following set of coupled equations:

$$\ddot{x}_a + \gamma_a \dot{x}_a + \omega_0^{a2} x_a + g x_b = F_a e^{i\omega t}, \quad (4)$$

$$\ddot{x}_b + \gamma_b \dot{x}_b + \omega_0^{b2} x_b + g x_a = F_b e^{i\omega t}, \quad (5)$$

where $x_{a,b}$ are the variable amplitudes of the oscillators $a$ and $b$, $g$ is the coupling between oscillators. The number of coupled modes can be more than 2, but for reasons of clarity only two coupled modes are considered in (4) and (5). By solving Eqs. (4) and (5) under the assumption of zero external force $F_{a,b} = 0$ and considering a wave solution $x_{a,b} \sim e^{i\omega t}$, one gets expressions for $\Omega_0^n, \Gamma_n, G_n$, as the functions of $\omega_0^a, \omega_0^b, \gamma_a, \gamma_b, g$ and the expression for the amplitude $A_n$ in the following form [21–24]:

$$|A_n| \approx A_0^n \left( G_n + 2\frac{\omega - \Omega_0^n}{\Gamma_n} \right) \bigg/ \sqrt{1 + 4\left(\frac{\omega - \Omega_0^n}{\Gamma_n}\right)^2}, \quad (6)$$

where $A_0^n$ is a constant which can be defined from either the initial conditions or can be normalized. Using a Tailor expansion and the same approach as previously outlined [the superposition of the solutions $S_{ij} = \sum A_n(\Omega_0^n, \Gamma_n, G_n, \omega)$ for a narrow frequency windows], the theoretical curve can be constructed. The function observed was similar to (3) but with differently defined terms:

$$S_{21} = A_1 + A_2\omega + \sum_{n=1}^{N} |S_{21\,\text{max}}|$$

$$\times \left( G_n + 2\frac{\omega - \Omega_0^n}{\Gamma_n} \right) \bigg/ \sqrt{1 + 4\left(\frac{\omega - \Omega_0^n}{\Gamma_n}\right)^2}. \quad (7)$$

In all the figures presented $f = \omega/2\pi$. The theoretical line in Fig. 8(a) matches $S_{ij}$ measurements well and the predicted positions of the eigenmodes (indicated by dots) are in good agreement with the maxima measured. In this case no additional eigenmodes were required to be added to match the measurements. However, Fig. 8(a) shows that there are still inaccuracies between experimental $S_{ij}$ and theoretical predictions. To improve the analysis further each group of coupled modes should be defined separately inside a frequency window, which should be as narrow as possible. As mentioned previously the expression (7) has been used as a continuous function throughout the whole frequency range, i.e. 1.27 to 1.305 GHz. To improve accuracy, the frequency windows for each group of the modes should be selected individually, and in Figs. 8(b)–8(e) the comparison between the theoretical and measured $S_{ij}$ curves inside a set of these intervals is shown. Each window was specially selected for a specific couplet or triplet group of modes. Interestingly only one mode located around 1.278 GHz is an uncoupled, single mode. By comparing the theoretical predictions with the measured dependences, it is clear that the fit with Fano-like solutions and the measurements are in a very good agreement. Another advantage of understanding the eigenmodes coupling is that it may be helpful for the mode control and cavity tuning. For example, by introducing a load for one mode in a specific group any other mode from the same group will also be loaded. In Fig. 9 the comparison of the cavity eigenmode spectrum is shown, $Q$-factors evaluated from the measured $S_{ij}$ (solid line) are compared with ones (empty circles) defined by 3D CST Microwave Studio. The fluctuation of the measured $S_{ij}$ (solid dots) is due to the uncertainty in the position of the couplers. In spite of this uncertainty a good agreement is achieved and the measurements of $S_{ij}$ were very stable and reliable.

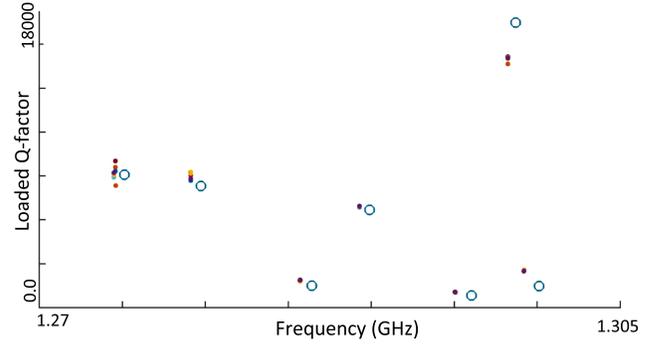

FIG. 9. Comparison of the $Q$-factors of the cavity PB eigenmodes calculated using CST MW Studio (circles) and evaluated from measurements (multiple dots) of the transmission parameters. The fluctuations of the measured frequency positions and amplitudes are an illustration of the uncertainty in the location of the rf couplers for each separate measurement.

## IV. STUDY OF EIGENMODES FIELD DISTRIBUTION USING RF BEAD-PULL TEST TABLE

To study the eigenmodes' field structure along the longitudinal coordinate the rf bead-pull test table has been used and the results have been compared with theoretical predictions. The schematic of the experiment is shown in Fig. 10(a) and can be seen in the photograph in Fig. 4. In the bead-pull measurements a small (as compared with the operating wavelength) dielectric spherical bead is moved slowly from port 2 to port 4. The bead interferes with the field inside the cavity and the strength of the interference is measured by measuring the $S_{13}$ parameter. The square of the reflected electric field is proportional to the relative shift of the eigenmode frequency, while the relative change in frequency is proportional to the tangent of the phase of $S_{21}$ [25–27] (the measurable parameter):

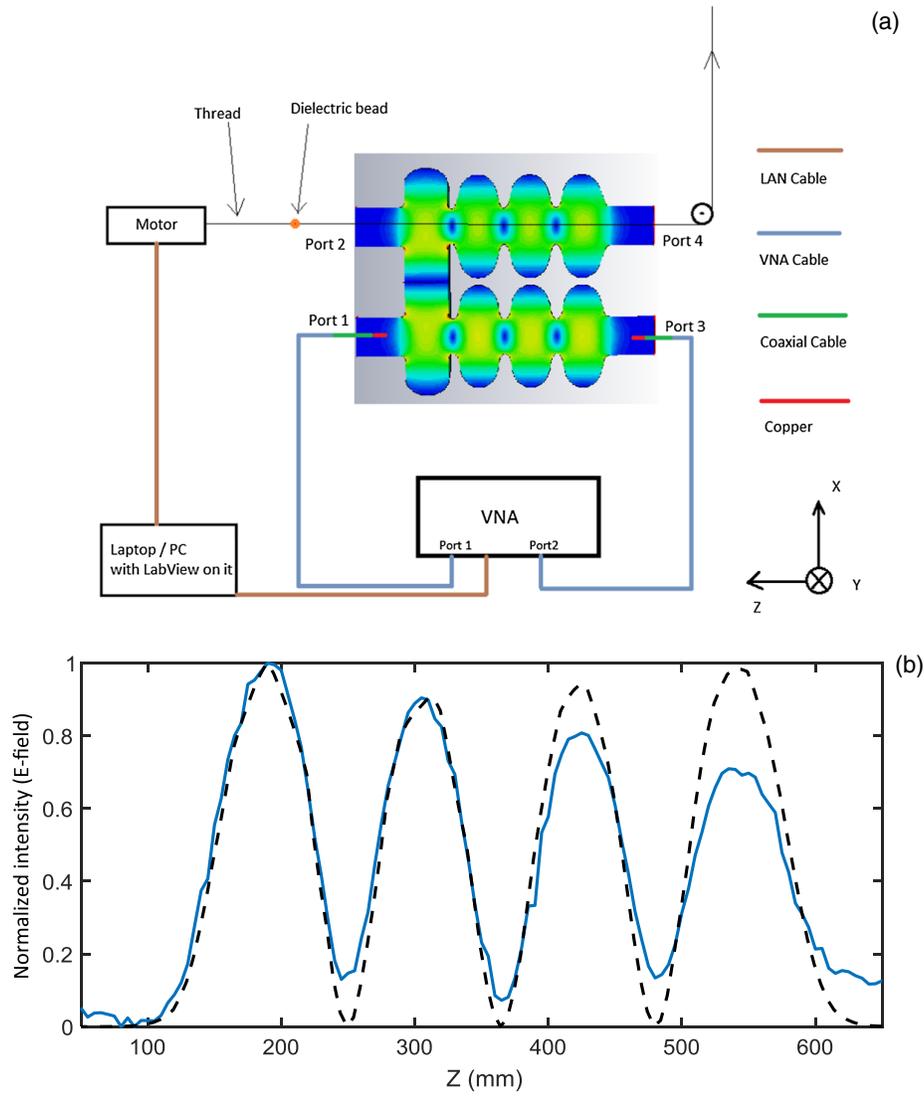

FIG. 10. (a) The schematic of the rf bead-pull measurements. (b) Comparison of the electric field normalized intensity distribution of cavity operating eigenmode at the frequency 1.2992299 GHz: calculated (dashed line) using CST MW Studio and measured (solid line) along cavity axis-2 (ports 4–2) while the field is excited along the axis-1 (ports 3–1) using dipole antennas.

$$\frac{\tan[\varphi(\omega_0^i)]}{2Q_L^i} = \frac{\omega_p^i - \omega_0^i}{\omega_0^i}, \quad (8a)$$

$$\frac{\omega_p^i - \omega_0^i}{\omega_0^i} = \frac{-\pi a^3 \epsilon_0(\epsilon_r - 1)}{U(\epsilon_r + 2)}(E_i)^2, \quad (8b)$$

where $\omega_{0,p}^i$ are the unperturbed and perturbed (subindex 0 and subindex $p$, respectively) eigenfrequencies of the "$i$"eigenmode, $a$ is the radius of the bead, $\epsilon_{0,r}$ is the dielectric permittivity of vacuum and the dielectric bead respectively, $\varphi$ is the phase measured during the experiments, $Q_L^i$ is the loaded $Q$-factor of the eigenmode (see Sec. III) and $U$ is the total energy stored in the cavity. It is clear that the field structure can be investigated using the expressions (8) via measurements of phase and if $\epsilon_r \gg 1$ the measurements can be very accurate. In our case the $\epsilon_r \sim 2.5$ and in Fig. 10(b) the profiles of the operating modes measured (solid line) are compared with theoretical prediction (dashed line). There are some deviations between measurements and numerical predictions as a result of the finite accuracies of the machining, eigenmode position definition and the measurement technique. As before, there are a number of different techniques which have been tested for measuring the field profiles and here only one, which we found to be accurate and reliable in this specific case of a two axis cavity is discussed. The method is as follows: first we set the bead-pull motor to move the bead by a small step of 5 mm and stop. At this stage the VNA is set to take a number of measurements in CW mode at a predefined frequency which corresponds to the frequency of the eigenmode $\omega_0^i$. The real and imaginary values of the measured $S_{21}$ parameter, the phase and the amplitude, are recorded and averaged out (over a number of measurements are taken while the probe is stationary). This method showed the best results when compared with other techniques. There are clear advantages of this method: making a stop at each step allows minimizing the vibration of the bead and averaging over a large number of measurements leads to a result with less deviation from the mean value. The procedure is fully computer controlled and all measurements were repeatable and consistent; automatization of the measurements led to noise level reduction as the measurements were done without anyone in the laboratory (no vibrations or air circulation). As a result the data observed was very close to the predictions (Fig. 11). In Fig. 11(a) comparison between measured (solid line) and predicted (dashed line) profiles (the normalized intensity of the field amplitudes) of the first seven eigenmodes is shown. The predicted profiles were calculated using CST MW Studio software. The "bridge mode" [Fig. 11(a)] is localized inside the coupling cell, as predicted, while the passband (PB) asymmetric mode which has highest $Q$-factor, has no field inside the coupling cell (fig. 11(c)). As expected, the modes with the highest $Q$-factor show better agreement between theory and experiment. This happens as the $Q$-factor of some of the eigenmodes is too small (around 1000) and coupling between the antenna and the mode is not efficient to excite them (the power of the VNA is not sufficient enough to overcome the excitation threshold value and the field amplitude inside the cavity becomes comparable with the noise). By simply increasing the input power the problem has been resolved for some of the modes, but also selection of the correct material for the dielectric bead with large $\epsilon_r \sim 100$ (TiO$_2$ for example) may further improve the measurements. An example of such a low-$Q$ mode is given by mode 1 [Fig. 7(b)]. To measure its structure the VNA input power was increased to the maximum level before the field amplitude dependence on the longitudinal coordinate has been measured, however, Fig. 11(a) shows that there are still some amplitude fluctuations with respect to theoretical predictions. In Fig. 12 the comparison of the measured profiles of the operating mode (solid line) and the PB modes having the highest $Q$-factor (broken lines) is shown.

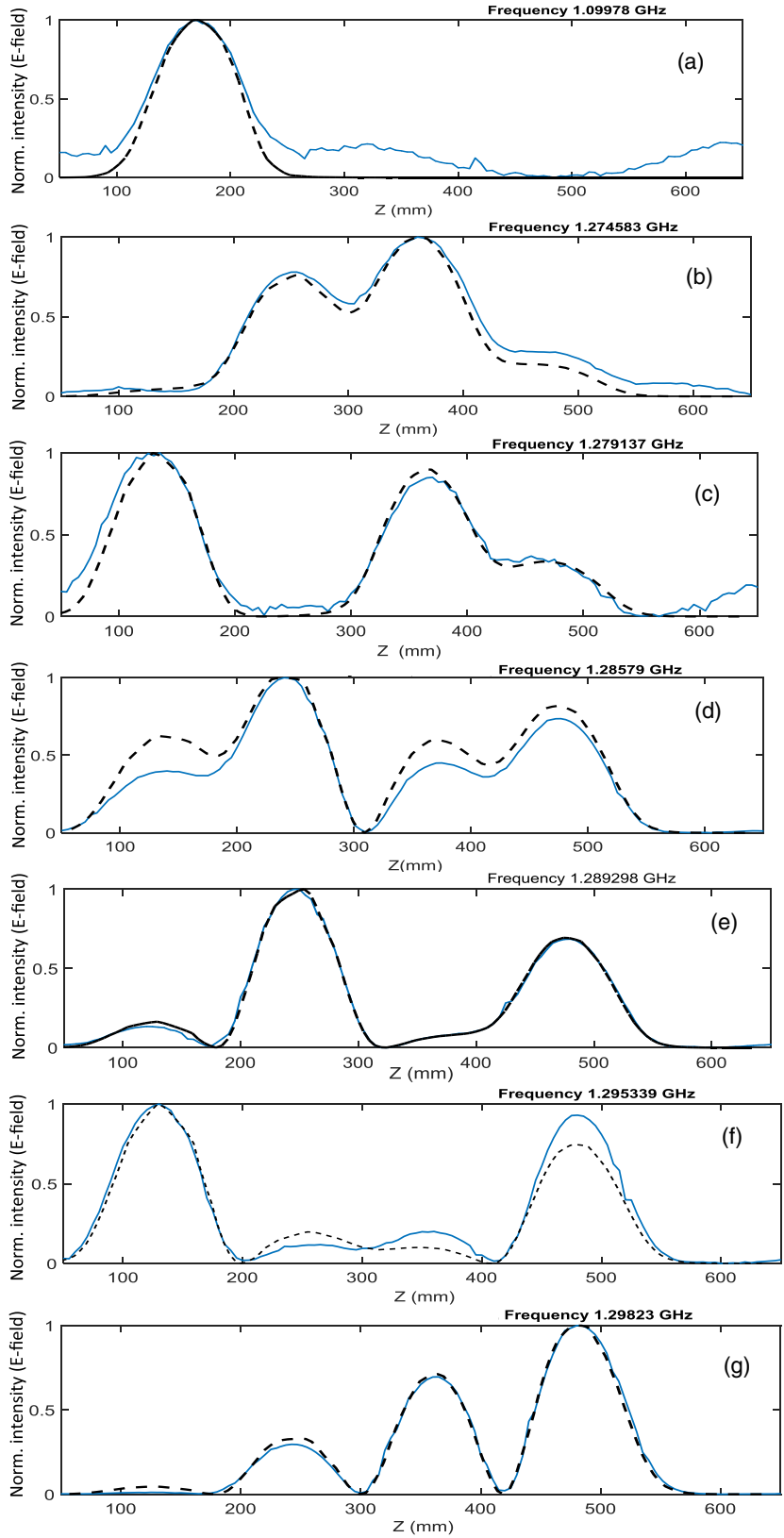

FIG. 11. Comparison of the normalized intensity distribution of the electric field of the cavity bridge and passband eigenmodes measure (solid line) as indicated in Fig. 8(a) and calculated (dashed line) at the frequencies: (a) 1.09978 GHz—bridge mode; (b) 1.274583 GHz—first passband mode; (c) 1.279137 GHz—second passband mode; (d) 1.28579 GHz—third passband mode; (e) 1.289298 GHz—fourth passband mode; (f) 1.295339 GHz—fifth passband mode; (g) 1.29823 GHz—sixth passband mode.

## V. CONCLUSION

In this paper a prototype of the aluminum dual axis 7-cell asymmetric rf cavity was studied. The cavity is a scaled down prototype of the 11-cell cavity designed for SCRF AERL to allow high-current operation. The numerical studies of the 7-cell cavity have been conducted using CST Microwave Studio and the results were discussed and compared with those observed during the previous studies. The cavity prototype has been machined from two blocks of aluminum using computer controlled equipment. This type of cavity machining was unconventional and proved to be cost effective, reliable and fast. A cavity consisting of two blocks can be easily assembled and disassembled (if new modifications are required), no soldering was needed, which made studying the prototype very fast and convenient. This cavity has been assembled and different techniques to measure the cavity's eigenmodes and their field profiles have been investigated. The first experimental studies of a dual axis 7-cell asymmetric rf cavity have been carried out and the results were presented and discussed. A good agreement between measured and predicted (3D CST MW Studio) eigenmode spectra was demonstrated. A novel coupled oscillators approach (Fano-like model) to describe the cavity eigenmodes was used and very good agreement between the measured and theoretical $S_{ij}$ curves was achieved. It was shown that in the cavity some eigenmodes are coupled and can be grouped together. Deeper understanding of mode grouping and coupling may improve the cavity tuning and control over the modes in the cavity. The results of spectral studies of the rf cavity have been compared with theoretical predictions and good matches have been demonstrated. The field structures of the eigenmodes located in the frequency range 1 to 1.5 GHz (encompassing passband modes and bridge modes), have been studied and presented. The results were compared with numerical simulations and discussed. Good agreement between predicted and measured field profiles for the modes with a high $Q$-factor has been demonstrated. The results observed are the first steps towards improvements of the measurement techniques described and ultimately toward the development of an asymmetric SCRF cavity for high average current ERL. Numerical modeling was demonstrated to be accurate even for these very complex cavities and its use will be essential to further improving the cavity design and the development of HOM loads and rf couplers. The next step will be the assembly of a full scale 11-cell copper cavity [Fig. 4(c)] to conduct rf tests and studies of a full set of modes including HOMs.

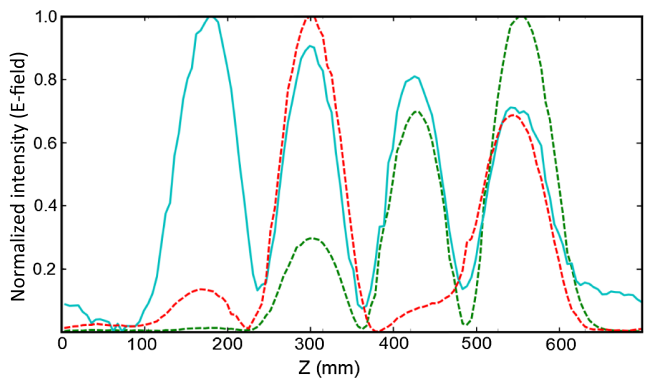

FIG. 12. Comparison of the normalized intensity distribution of the electric fields of the cavity passband eigenmodes as in Fig. 10 (a): operating (solid line) located at 1.299 229 9 and HOMs located at 1.298 23 GHz (green dashed line) and 1.289 298 GHz (red dotted line).

## ACKNOWLEDGMENTS

The rf-test laboratory and equipment were provided by John Adams Institute (Oxford) supported by the U.K. STFC, Grant No. ST/J002011/1, and the authors would like to thank JAI Oxford for the partial support of the project. Ivan Konoplev and Andrew Lancaster would like to thank STFC U.K. and The Leverhulme Trust for the partial support via PRD grant (ST/M003590/1) and Network Grant (IN 2015 012) which provided the means for software license and travel. Kaloyan Metodiev would like to thank the Department of Physics, University of Oxford, and JAI Oxford for support of his summer research project. Ivan Konoplev would like to thank Ms. Hannah Harrison for helping with the text editing.

[1] M. Tonouchi, Cutting-edge terahertz technology, Nat. Photonics 1, 97 (2007).

[2] H.-T. Chen, W. J. Padilla, J. M. O. Zide, A. C. Gossard, A. J. Taylor, and R. D. Averitt, Active terahertz metamaterial devices, Nature (London) 444, 597 (2006).

[3] J. F. Federici, B. Schulkin, F. Huang, D. Gary, R. Barat, F. Oliveira, and D. Zimdars, THz imaging and sensing for security applications—explosives, weapons and drugs, Semicond. Sci. Technol. 20, S266 (2005).

[4] G. L. Carr, M. C. Martin, W. R. McKinney, K. Jordan, G. R. Neil, and G. P. Williams, High-power terahertz radiation from relativistic electrons, Nature (London) 420, 153 (2002).

[5] V. Cnudde and M. N. Boone, High-resolution x-ray computed tomography in geosciences: A review of the current technology and applications, Earth-Sci. Rev. 123, 1 (2013).

[6] C. Clavero, Plasmon-induced hot-electron generation at nanoparticle/metal-oxide interfaces for photovoltaic and photocatalytic devices, Nat. Photonics 8, 95 (2014).

[7] G. H. Hoffstaetter and I. V. Bazarov, Beam-breakup instability theory for energy recovery linacs, Phys. Rev. ST Accel. Beams 7, 054401 (2004).

[8] T. Nakamura, Multiturn circulation of an energy-recovery linac beam in a storage ring, Phys. Rev. ST Accel. Beams 11, 032803 (2008).

[9] D. W. Feldman et al., in Proceedings of the 1987 Particle Accelerator Conference (IEEE, Washington, DC, 1987), p. 221; Energy recovery in the Los Alamos free electron laser, Nucl. Instrum. Methods Phys. Res., Sect. A 259, 26 (1987).

[10] C.-X. Wang, Conceptual design considerations of a 5-cell dual-axis SRF cavity for ERLs, in Proceedings of SRF2007 (Peking University, Beijing, China, 2007).

[11] R. Ainsworth, G. Burt, I. V. Konoplev, and A. Seryi, Asymmetric dual axis energy recovery linac for ultrahigh flux sources of coherent x-ray and THz radiation: Investigations towards its ultimate performance, Phys. Rev. Accel. Beams 19, 083502 (2016).

[12] A. Seryi, Ultrahigh flux compact x-ray source, Patent PCT Application No. PCT/GB2012/052632 (24 Oct 2012).

[13] R. Ainsworth, G. Burt, I. V. Konoplev, and A. Seryi, Asymmetric superconducting rf structure, Patent PCT Application No. PCT/GB2015/053565 (24 Nov 2015).

[14] I. V. Konoplev, A. Seryi, A. J. Lancaster, K. Metodiev, G. Burt, and R. Ainsworth, Compact, energy efficient superconducting asymmetric ERL for ultrahigh fluxes of x-ray and THz, AIP Conf. Proc. 1812, 100004 (2017).

[15] B. Aune et al., The superconducting TESLA cavities, Phys. Rev. ST Accel. Beams 3, 092001 (2000).

[16] R. Ainsworth, G. Boorman, A. Lyapin, S. Molloy, A. Garbayo, P. Savage, A. P. Letchford, and C. Gabor, Bead-pull test bench for studying accelerating structures at RHUL, in Proceedings of the 2nd International Particle Accelerator Conference, San Sebastián, Spain (EPS-AG, Spain, 2011), MOPC049.

[17] D. Kajfez and E. J. Hwan, Q-factor measurement with network analyzer, IEEE Trans. Microwave Theory Tech. 32, 666 (1984).

[18] P. J. Petersan and S. M. Anlage, Measurement of resonant frequency and quality factor of microwave resonators: Comparison of methods, J. Appl. Phys. 84, 3392 (1998).

[19] W. Xu et al., Report No. BNL-98926-2012-IR, 2012.

[20] F. Furuta, R. G. Eichhorn, G. M. Ge, D. Gonnella, G. H. Hoffstaetter, M. Liepe, P. Quigley, and V. Veshcherevich, HOM measurements for Cornell's high-current CW ERL cryomodule, in Proceedings of IPAC2016, Busan, Korea, 2016, WEPMR021 (JACoW, Geneva, 2016), pp. 2309–2311.

[21] U. Fano, Effects of configuration interaction on intensities and phase shifts, Phys. Rev. 124, 1866 (1961).

[22] R. Kumar, Asymmetry to symmetry transition of Fano line shape: Analytical derivation, Indian J. Phys. 87, 49 (2013).

[23] G. Vaisman, E. O. Kamenetskii, and R. Shavit, Magnetic-dipolar-mode Fano resonances for microwave spectroscopy of high absorption matter, J. Phys. D 48, 115003 (2015).

[24] A. E. Miroshnichenko, S. Flach, and Y. S. Kivshar, Fano resonances in nanoscale structures, Rev. Mod. Phys. 82, 2257 (2010).

[25] J. Guillaume, CERN Summary Report, 2015, https://cds.cern.ch/record/2044542/files/G_Jaume_SUMM_Report.pdf.

[26] M. Navarro-Tapia and R. Calaga, Bead-pull measurements of the main deflecting mode of the double-quarter-wave cavity for the HL-LHC, in Proceedings of SRF2015, Whistler, BC, Canada, 2015, THPB019 (JACoW, 2015), pp. 1105–1109.

[27] H. Hahn, W. Xu, P. Jain, and E. C. Johnson, HOM identification and bead pulling in the Brookhaven ERL, in Proceedings of SRF2011, Chicago, IL, 2011, THPO041, pp. 810–817.